\documentclass[runningheads]{svmult}

\usepackage{makeidx}   % allows index generation
\usepackage{graphicx}  % standard LaTeX graphics tool
\usepackage{subeqnar}  % subnumbers individual equations
\usepackage{multicol}  % used for the two-column index
\usepackage{physprbb}  % modified textarea for proceedings,
\makeindex             % used for the subject index
\begin{document}
\title*{Intergalactic Magnetic Fields 
\protect\newline from Quasar Outflows}
\toctitle{Intergalactic Magnetic Fields
\protect\newline  from Quasar Outflows}
% allows explicit linebreak for the table of content
%
%
\titlerunning{Intergalactic Magnetic Fields}
% allows abbreviation of title, if the full title is too long
% to fit in the running head
%
\author{Steven R. Furlanetto
\and Abraham Loeb}
%
%\authorrunning{Ivar Ekeland et al.}
% if there are more than two authors,
% please abbreviate author list for running head
%
%
\institute{Department of Astronomy, Harvard University, Cambridge MA
02138, USA}

\maketitle              % typesets the title of the contribution

\begin{abstract}
Outflows from quasars inevitably pollute the intergalactic medium (IGM)
with magnetic fields. The short-lived activity of a quasar leaves behind an
expanding magnetized bubble in the IGM. We model the expansion of the
remnant quasar bubbles and calculate their distribution as a function
magnetic field strength at different redshifts.  We 
find that by a redshift $z\sim 3$, about 5--$80\%$ of the IGM volume is
filled by magnetic fields with an energy density $> 10\%$ of the mean
thermal energy density of a photo-ionized IGM (at $\sim 10^4$ K).  As
massive galaxies and X-ray clusters condense out of the magnetized
IGM, the adiabatic compression of the magnetic field could result in
the fields observed in these systems without a need for further dynamo
amplification.  
\end{abstract}

\section{Introduction}

Clusters of galaxies contain substantial magnetic fields
with strengths $B \sim 0.1-10\;\mu$G and coherence lengths $\ell
\sim 10$ kpc~\cite{furlanetto:kronberg}.  The origin of such fields
could have important implications for structure formation.  Assuming flux
conservation, a cluster field $B_{\rm cl} \sim 10^{-7}$~G
would imply $B_{\rm IGM} \sim 10^{-9}$~G in 
the diffuse intergalactic medium (IGM), which would constitute $\sim
5\%$ of the 
thermal energy density of a photoionized IGM.  The observational
constraints on an intergalactic magnetic field (IGMF) are weak,
requiring only that $B_{\rm IGM} < 10^{-8}(\ell/\mbox{
Mpc})^{-1/2}$~G in the currently popular 
$\Lambda$CDM model~\cite{furlanetto:kronberg}.

Unfortunately, there is no convincing model for the formation of cluster
fields or the
IGMF.  Because the rotation times of clusters exceed the Hubble time,
dynamos are expected to be ineffective.  Primordial generation
scenarios cannot explain the large coherence
lengths~\cite{furlanetto:quashnock}.  Models that generate magnetic
fields during structure formation~\cite{furlanetto:kulsrud} or in
starbursts~\cite{furlanetto:kronberg-lesch} suffer from similar problems.  

We have examined the possibility that the IGMF was originally produced
near supermassive black holes and expelled into the IGM through mechanical
outflows~\cite{furlanetto:rees} from radio-loud quasars (RLQs) and broad
absorption line quasars (BALQs).  Supermassive black holes are
one of the few classes of astrophysical objects with energy
reservoirs large enough to account for the large-scale
fields in clusters, and the relatively small number of
powerful sources can accomodate the large observed coherence lengths,
as emphasized by~\cite{furlanetto:colgate}.
For more details on the model, see~\cite{furlanetto:furl}.

\section{ Filling Factor of Magnetized Regions \& Mean Fields }

While a quasar is active, its outflow is in the form of twin
collimated jets (for a RLQ) or equatorial winds (for a BALQ).  After
the quasar becomes dormant, the outflow remnant is 
overpressured with respect to the IGM and continues to
expand in comoving coordinates until its outward velocity matches the
Hubble flow velocity.  We assume that the outflow remnant isotropizes
and expands adiabatically as a spherical shell during this late phase.

Simple energy conservation 
provides a surprisingly accurate estimate of the final comoving bubble
size $\hat{R}_{\rm max}$.  Balancing the energy input of the quasar
and the final kinetic energy of the shell, we find that
$\hat{R}_{\rm max} \propto [L_q \tau_q \varepsilon_K (1 +
\varepsilon_B)]^{1/5}$,
where $L_q$ is the luminosity of the quasar, $\tau_q$ is
its lifetime, $\varepsilon_K$ is the ratio of the mechanical and radiative
luminosities, and $\varepsilon_B$ is the ratio of the
magnetic and mechanical energy outputs.  Note the weak dependence on
the quasar parameters, making our results robust to uncertainties in
their measurement.  In particular, we find that magnetic fields do not play
an important role in the expansion for realistic values of
$\varepsilon_B$.  We therefore ignore the geometric and
magnetohydrodynamic effects of the magnetic field and assume simple
flux conservation.  

We next calculate the number of quasar sources.  For $z < 4$, we
use the observed optical luminosity function of
quasars~\cite{furlanetto:pei} together with 
an assumed incidence rate of outflows $f$.  For $z > 4$, we assume
that the incidence rate of quasars is proportional to the
Press-Schechter mass function, with the proportionality constant set
by matching to the observed luminosity function at $z \sim
4$~\cite{furlanetto:haiman}.  

We examine two quasar models: the RLQ model, with $\varepsilon_B =
0.1$ and $f = 0.1$~\cite{furlanetto:stern}, and the BAL model, with
$\varepsilon_B=0.01$ and $f=1$~\cite{furlanetto:weymann}.  Note that
there are no existing observations of magnetic fields in BALQs, though
the fields may play an important role in the wind
mechanism~\cite{furlanetto:dekool}. 

We then calculate the total filling factor of the quasar bubbles as a
function of redshift, $F(z)$, and the global volume-averaged magnetic
energy density, $\bar{u}_B(z)$, for our two models by numerically
integrating the equation of motion of each remnant and summing over
all quasar sources.  Our results are shown in Fig. 1, along with the
distribution of magnetic field strengths for a series of redshifts.  

\begin{figure}[b]
\begin{center}
\includegraphics[width=5cm]{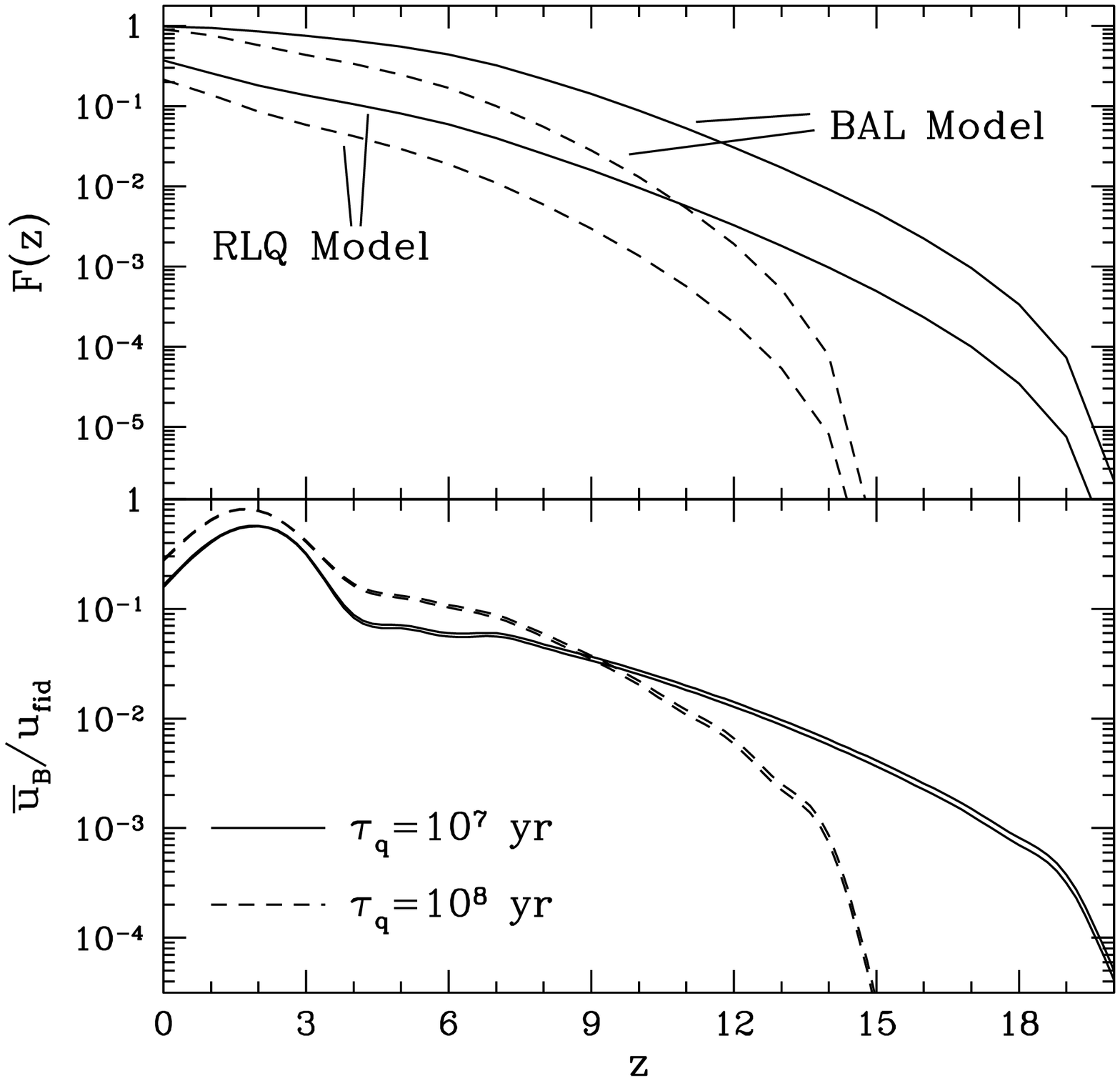}
\includegraphics[width=5cm]{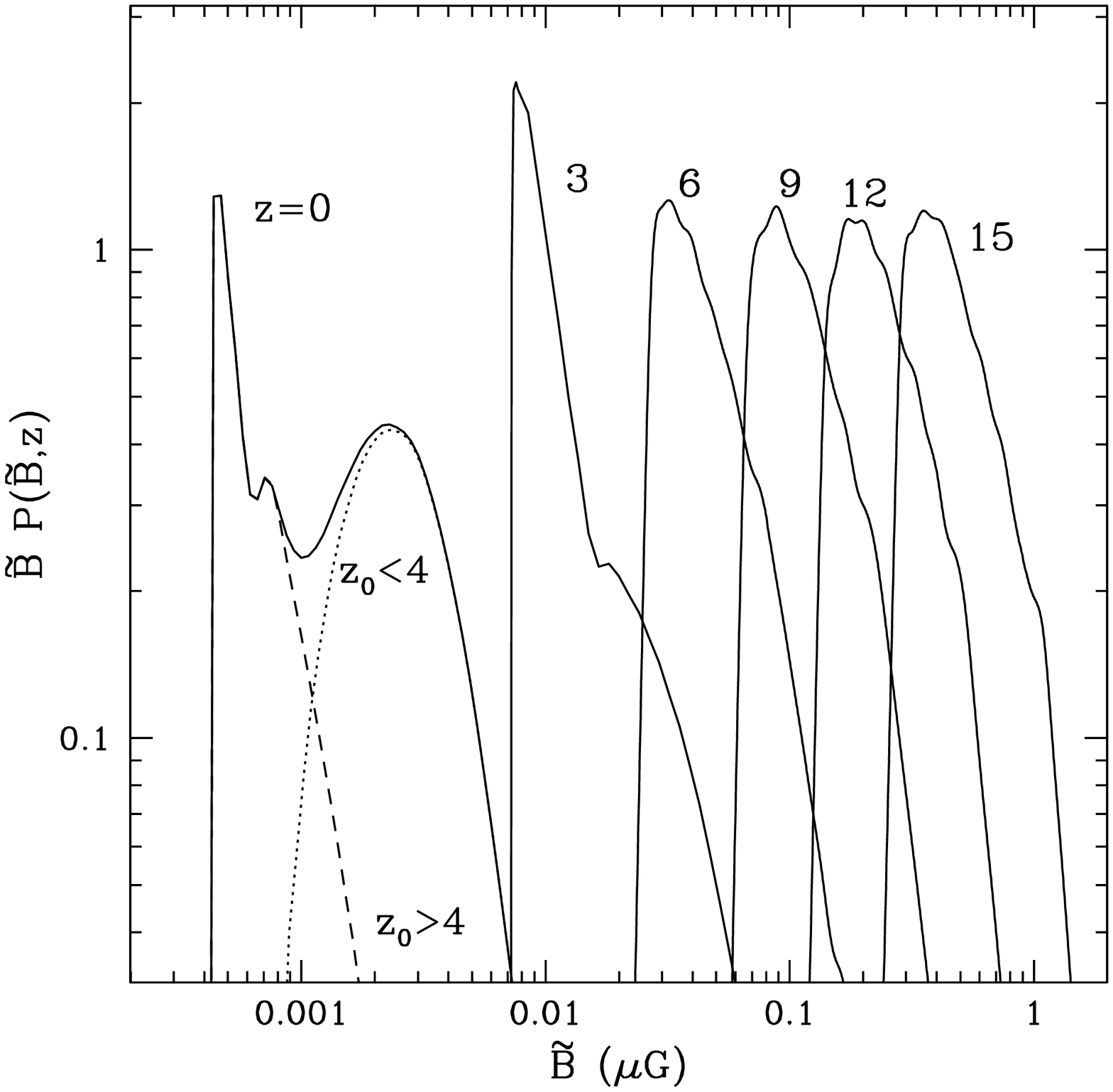}
\end{center}
\caption{ \emph{Top Left:} Volume filling fraction of magnetized bubbles
$F(z)$.  \emph{Bottom Left:}
Ratio of magnetic energy density, $\bar{u}_B$, to the fiducial thermal
energy density $u_{fid} = 3 n(z) k T_{IGM}$, where 
$T_{IGM} = 10^4$ K.  In each of these panels, results are shown for
both the RLQ and BALQ models.
\emph{Right:} Probability distributions of bubble magnetic field,
$P(\tilde{B},z)$, for the RLQ model, at various redshifts.  Here
$\tilde{B} = B/(\sqrt{\epsilon_B/0.1})$; in these units the curves are
 independent of $\epsilon_B$.  The dotted curve shows the contribution
to the $z=0$ distribution function from quasars forming at $z_0 < 4$
and the dashed curve shows the contribution from quasars at $z_0 > 4$.
All curves are normalized to have unit area. }
\label{fig1}
\end{figure}

\section{Results} 

We predict a cellular IGMF filling a substantial fraction of space by
$z \sim 3$ (Fig. 1), with each cell a fossil bubble produced by
a single magnetized quasar outflow.  Cells generated by RLQ
quasars are more highly magnetized than those from BAL
quasars but fill a smaller volume.  We
predict that $B_{\rm IGM} \sim 10 f \varepsilon_B$~nG when averaged
over all of space; for each of our models, simple adiabatic compression of
such an IGMF can account for the observed cluster magnetic fields.

Direct detection of the cells via Faraday rotation measurements will
be difficult, but electron acceleration by shocks in the cells will
cause synchrotron emission.  The same accelerated electrons
produce $\gamma$-rays through inverse-Compton scattering of the cosmic
microwave background.  Correlated maps of the $\gamma$-ray
and radio skies may allow us to calibrate the magnetic field in the
shocks~\cite{furlanetto:waxman}.  

Finally, we note that the non-thermal pressure of the magnetic fields
may help to resolve the discrepancy between simulated and observed
line widths in the Ly$\alpha$ forest~\cite{furlanetto:bryan}.

%\begin{figure}[htbp]
%\special{psfile=furlanettosf1.eps hoffset=17 voffset=-190 hscale=25
%vscale=25} 
%\special{psfile=furlanettosf2.eps hoffset=177 voffset=-190 hscale=25
%vscale=25} 
%\vspace{2.2in}
%\caption{ \emph{Top Left:} Volume filling fraction of magnetized bubbles
%$F(z)$.  \emph{Bottom Left:}
%Ratio of magnetic energy density, $\bar{u}_B$, to the fiducial thermal
%energy density $u_{fid} = 3 n(z) k T_{IGM}$, where 
%$T_{IGM} = 10^4$ K.  In each of these panels, results are shown for
%both the RLQ and BALQ models.
%\emph{Right:} Probability distributions of bubble magnetic field,
%$P(\tilde{B},z)$, for the RLQ model, at various redshifts.  Here
%$\tilde{B} = B/(\sqrt{\epsilon_B/0.1})$; in these units the curves are
%independent of $\epsilon_B$.  The dotted curve shows the contribution
%to the $z=0$ distribution function from quasars forming at $z_0 < 4$
%and the dashed curve shows the contribution from quasars at $z_0 > 4$.
%All curves are normalized to have unit area. }
%\label{fig1}
%\end{figure}

%INDEX%%%%%%%%%%%%%%%%%%%%%%%%%%%%%%%%%%%%%%%%%%%%%%%%%%%%%%%%%%%%%%%
% Please check with the editor of your book whether he plans to
% include a "mutual" subject index - if so, please code your entries
% in the standard syntax. For your own purposes you may print your
% "personal" index by using the following commands:
%
%\clearpage
%\addcontentsline{toc}{section}{Index}
%\flushbottom
%\printindex
%%%%%%%%%%%%%%%%%%%%%%%%%%%%%%%%%%%%%%%%%%%%%%%%%%%%%%%%%%%%%%%%%%%%%

\end{document}